\documentclass[aps,prl, twocolumn,superscriptaddress]{revtex4}
\usepackage{graphicx}
\usepackage{amsmath}
\usepackage{amsfonts}
\usepackage{amsthm}
\usepackage{amssymb}
\usepackage{amsbsy}
\usepackage{pstricks}
\usepackage{wasysym}
\usepackage{mathrsfs}

\input{epsf}
\newcommand{\beq}{\begin{equation}}
\newcommand{\eeq}{\end{equation}}
\newcommand{\bea}{\begin{eqnarray}}
\newcommand{\eea}{\end{eqnarray}}

\newcommand{\ef}{\epsilon_F}
\newcommand{\ev}{E_v}
\newcommand{\flg}{F_{LG}}
\newcommand{\rhopf}{\rho_{\text{\tiny Pf}}}

\def\br{\mathbf{r}}

\begin{document}

\title{The Weakly Coupled Pfaffian as a Type I Quantum Hall Liquid}
\author {S. A. Parameswaran}
\email{sashok@princeton.edu}
\affiliation{Department of Physics, Princeton University, Princeton, NJ 08544}
\author {S. A. Kivelson}
\email{kivelson@stanford.edu}
\affiliation{Department of Physics, Stanford University, Stanford, CA 94305}
\author {S. L. Sondhi}
\email{sondhi@princeton.edu}
\affiliation{Department of Physics, Princeton University, Princeton, NJ 08544}
\author {B.Z. Spivak}
\email{spivak@u.washington.edu}
\affiliation{Department of Physics, University of Washington, Seattle, WA 98195}

\date{\today}
\begin{abstract}

The Pfaffian phase of electrons in the proximity of a half-filled Landau level is understood to
be a $p+ip$ superconductor of composite fermions.
We consider the properties of this paired quantum Hall phase when
the pairing scale is small, {\it i.e.} in the
weak-coupling, BCS, limit, where the coherence length is  much larger than the charge screening length.
We find that, as in a Type I superconductor, the vortices attract so that, upon
varying the magnetic field from its magic value at $\nu=5/2$, the
system exhibits Coulomb frustrated phase separation. We propose that the weakly and strongly
coupled Pfaffian states exemplify a general dichotomy between Type I and Type II quantum Hall
fluids.
\end{abstract}
\maketitle

\addtolength{\abovedisplayskip}{-1mm}

\noindent
There is a deep and precise relation between superconductivity and the quantum Hall effect, which can be formally implemented by replacing the physical Maxwell gauge field by the statistical Chern-Simons gauge field \cite{Girvin:1987p1415, ZHANG:1989p1367, Read:1989p1421}.  Here, there is a correspondence between the perfect conductivity of the superconductor and the quantization of the Hall conductance, the Meissner effect and incompressibility, and quantized vortices and fractionally charged quasiparticles.  In some cases, this relation goes even further, in that the Hall fluid is a condensate of electron pairs.  Specifically,
the quantized Hall phase seen in the proximity of the half-filled second Landau level, $\nu=5/2$,
is likely associated with the Pfaffian or Moore-Read state \cite{Moore:1991p58}.
The ideal Pfaffian state has a natural interpretation as a weakly-paired state of composite fermions \cite{Read:2000p1015, GREITER:1992p1262} with
$p+ip$ symmetry, {\it i.e.} it bears an analogous relationship to the metallic
``composite Fermi liquid'' state of the half-filled Landau level  \cite{Halperin:1993p144}  as the corresponding BCS state does to an ordinary metal.

In this letter we will explore an important aspect of the quantum-Hall/superconductor relationship, particularly relevant to the Pfaffian phase, which has not been explored fully in previous work.  Specifically, the relationship implies that quantum Hall
states will, like superconductors, generically exhibit two length scales: a screening length $\lambda$ that characterizes the
decay of density deviations and a coherence length $\xi$ \cite{Lu:2010p167} that characterizes variations of the superfluid order. Further, the ratio of these scales should crucially influence the structure of vortices and and thence the response of the ideal quantum Hall states to global density changes
much as they influence the response of superconductors to applied magnetic fields.

Accordingly we propose that quantum Hall fluids should come in two classes: a) Type II quantum Hall fluids where (roughly) $\lambda \geq \xi$ and density deviations are accommodated by the introduction of single quasiparticles/vortices and b) Type I quantum Hall fluids where $\lambda < \xi$  and quasiparticles are unstable to agglomeration and form multi-particle bound states or if the interactions are sufficiently short-ranged, phase separate entirely. Intuitively, Type I behavior arises at $\xi \gg \lambda $ as two vortices of size
$\xi$ are able to reduce their joint energy by an amount of order of their individual creation energies $E_v \sim (\xi)^0$ by merging and thus reducing the region over which the order parameter is suppressed while only paying an interaction which is parametrically smaller, e.g. a Coulombic cost
of order $e^2/\xi$.
\begin{figure}
\includegraphics[width = \columnwidth]{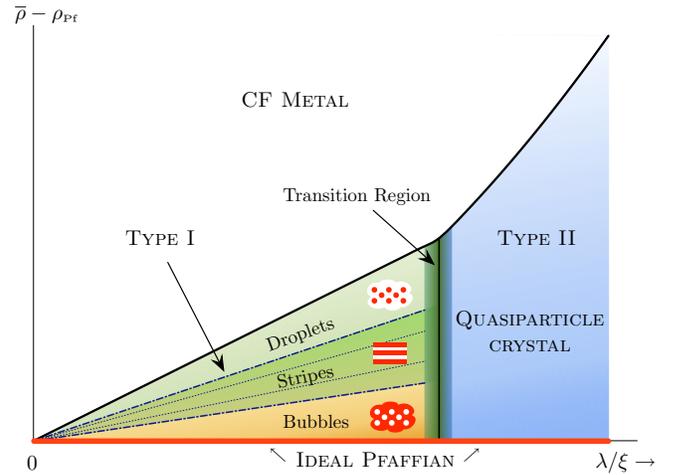}
\caption{\label{fig-phasediagram} Schematic phase diagram of the Pfaffian phases with  fixed Coulomb interactions as a function of density deviation from $\nu=5/2$ (pseudomagnetic field) and the Landau-Ginzburg parameter, $\lambda/\xi$.  Typical configurations of the different inhomogeneous phases are shown, with red representing the Pfaffian and white the metallic phase. For short-ranged interactions, the microemulsion phases in the Type I region are replaced by a phase-separated intermediate state, with the Pfaffian fraction decreasing continuously to zero as the boundary to the composite fermion metal is approached.}\end{figure}

Examples of Type II QH liquids abound---indeed, it has been implicitly assumed that {\it all} QH fluids are of that kind and that they exhibit a single length scale, which ultimately is the magnetic length ($\ell_B = \sqrt{\hbar c/eB}$). In this paper we will present and analyze a first example of a Type I QH liquid.

This liquid is the weak coupling limit of the Pfaffian state. On general grounds, the identification of the Pfaffian phase as a paired state of composite fermions carries with it a natural presumption that the pairing strength is tunable and that, in particular, it is possible to reach a weak coupling regime where, as in BCS superconductors, $\xi$ is parametrically larger than the zero-temperature charge screening length, $\lambda(T=0) \sim \ell_B$. Indeed, there is considerable evidence from numerics that such a weak coupling regime does indeed exist for suitable microscopic  interactions \cite{Rezayi:2000p1458,Moeller:2008p1073}. In the balance of this letter we develop a theoretical
framework in which to analyze the properties of the weakly coupled Pfaffian and use it to show
that instead of the various vortex/quasiparticle glass or crystal phases typically invoked for the quantum Hall plateaux, it exhibits patterns of charge organization associated with
frustrated phase separation.  A schematic phase diagram which summarizes the findings of the present study is shown in Fig. \ref{fig-phasediagram}.

\noindent{\bf  Landau-Ginzburg theory:} For concreteness, let us begin by thinking about the half-filled
Landau level within the framework of the fermionic Chern-Simons (CS) approach to the problem \cite{Lopez:1991p1}, in which two
 quanta of statistical flux are bound to each electron. Exactly at $\nu=5/2$,
if we treat the CS gauge field at mean-field level, the electrons form a Fermi sea of composite fermions in zero net field \cite{Halperin:1993p144}.  Following the lead of previous studies \cite{Foster:2003p750}, we add an explicit $p-$wave attraction between the composite fermions (which in a more complete treatment should properly be derived by integrating out high energy degrees of freedom in some fashion).
At $T=0$, a further BCS mean-field decoupling of this attraction produces a $p+ip$ superconducting state with coherence length $\xi$ and a gap to creating Bogoliubov quasiparticles (`Bogolons') $\Delta = \hbar v_F/\xi$.
Due to the Chern-Simons electrodynamics of the statistical gauge field, vortices carry charge $e/4$ which, as we will see shortly, is localized on a length scale $\lambda$ which plays the role of the London penetration depth in superconductors.  The Bogolons are neutral fermionic excitations of the quantum Hall fluid \cite{Moore:1991p58,Baraban:2009p1372,  Bonderson:2010p340,Moller:2010p1370}.

If the attractive interaction is weak, then $\Delta \ll \ef$ or equivalently $\xi \gg \lambda$. Moreover, in this limit, the vortex creation energy, $\ev \gg \Delta$.  (A standard estimate  -- which we will improve below -- yields $\ev \sim \ef$.) Thus, for weak coupling there exists a finite temperature regime where vortices are sparse, $n_v \sim e^{-\ev/T}$ and can be neglected on length scales $\ll\ell_B e^{+\ev/2T}$. There is a well-defined crossover  temperature $T_c^{MF}\sim \Delta/2$  corresponding to the mean-field transition, at which significant pairing onsets.

 Near $T_c^{MF}$, one can derive a Landau-Ginzburg description from the fermionic CS path integral by
integrating out the fermions. As we enforce a particular chirality of the $p$-wave pairing and
assume the electrons are spin-polarized, this will depend on a single complex order parameter $\Psi$.
To simplify the expression for $\flg$, we choose units such that in the uniform Pfaffian phase, the value of the order parameter is $|\Psi|=1$ and the condensation energy per unit area is $\epsilon_0/\xi^2$ (where $\epsilon_0 \sim \epsilon_F$). This yields:
\begin{widetext}\begin{eqnarray}\label{eq-LG}\frac{\flg} {\epsilon_0} &=& \int  {d\br}  \, \left[ \left| \left\{\frac{\nabla}{i} - \mathfrak{a}\right\}\Psi\right|^2 + \frac{1}{2\xi^2}\left(1 - \left|\Psi\right|^2\right)^2\right]  +{\lambda^2}
  \int d\br\,\left[\nabla \times \mathfrak{a}\right]^2 \nonumber\\ & &+  \frac{1}{2 }\int d\br d\br' \left[ \nabla\times\mathfrak{a} -8\pi\overline{\delta\rho}\right]_{\br}\left[\frac{\lambda_{C}}{|\br -\br'|}\right]^x \left[ \nabla\times\mathfrak{a} -8\pi\overline{\delta\rho}\right]_{\br'} + \int  {d\br}\,\frac{e}{8\pi}{A}_{0,\text{ext}}[\nabla\times\mathfrak{a}]
\end{eqnarray}
\end{widetext}
Here, the first term  is the usual expression for the energy of the condensed phase in powers of $\Psi$.  In the second term, which is formally equivalent to  a Maxwell term, $\lambda^2$ is largely proportional to the inverse compressibility of the  the HLR metal.  The gauge field $\mathfrak{a}(\br) \equiv \frac{8\pi}{\kappa}(\mathbf{a}(\br) +\mathbf{A}_{\text{ext}})$ is the sum of the Chern-Simons and external gauge fields, which satisfy $\nabla\times \mathbf{a} = \kappa\rho(\br)$, $\nabla\times \mathbf{A}_{\text{ext}} = -\kappa\rhopf=B$  respectively, where $\kappa^{-1} =e/2hc$ is the Chern-Simons coupling, and $\rhopf$ is the density of the half-filled Landau level of the ideal Pfaffian state. The third term, which has no direct analogue in the theory of superconductivity, is the Coulomb interaction, with $\overline{\delta\rho} = \bar\rho -\rhopf$ the difference between the density set by the positive background and the commensurate density. We have introduced an exponent $x$ which takes value $x=1$ when the Coulomb interaction is unscreened, but can take the value $x=3$ (corresponding to dipolar interactions) when there is a metallic gate present.  (For $x=1$, $\lambda_C \sim e^2/\epsilon_F$, while for the dipolar case, $\lambda_C^3 \sim e^2 d^2/\epsilon_F$ where $d$ is the distance to the metallic gate.)
 The final term ensures that the uniform superconducting (Pfaffian) state has the correct Hall response. $\xi$ and $\lambda$ represent the Landau-Ginzburg coherence length and penetration depth of the CS superconductor.

Two comments are in order. First, the formal derivation of the Landau-Ginzburg theory is valid only near $T_c^{MF}$ and at
lower temperatures non-local ``Pippard'' effects will need to be included as is the case for
Type I superconductors \footnote{At $T=0$ a self-consistent treatment of nonlocal effects yields an actual penetration depth $\sim (\lambda^2 \xi)^{1/3}$; see Ref.~\onlinecite{Tinkham:1996p1}. For the Pfaffian, this ensures that a sensible result obtains in the metallic limit as $\xi\rightarrow 0$.}. However, for a first pass at the problem, we will ignore the subtleties involved and use the Landau-Ginzburg theory
down to $T=0$ with its parameters considered to be phenomenological constants. Second, as $\flg$ is formally obtained by integrating out the composite fermions, the full partition function would be obtained by performing a further path-integral over the Hubbard Stratonovich field, $\Psi$, and the CS gauge field, $\mathbf{a}$. For present purposes, we will treat this problem in saddle-point approximation, where we focus on the configuration of $\Psi$ and $\mathbf{a}$ which minimizes $\flg$. This will be adequate for the purposes
of understanding vortex structure and interaction which is our focus in the remaining.

\noindent{\bf Vortex structure:}  We turn now to the vortices. First, let us consider the structure of the vortices in the absence of a long ranged interaction, ($\lambda_C=0$). Now $\flg$ has exactly the form of the conventional free energy of a superconductor, with screening length $\lambda$.
In the extreme Type I limit, $\lambda\ll \xi$, the vortex has a core with magnetic flux  spread roughly over a region of size $\lambda$ while the order parameter is suppressed over a much larger region of size $\xi$. Translated into quantum Hall language, the charge density of the quasiparticle is confined primarily to a region of size $\lambda$ which is much smaller than the region over which the Pfaffian order is disrupted.

This structure can be captured by a simple variational ansatz (for a vortex of vorticity $N$)
with a single parameter, $L$, that has been shown \cite{Yung:1999p1547} to be accurate in the limit $\xi/\lambda \rightarrow \infty$:
\begin{eqnarray}
\Psi(\br)  &=&  e^{i N\theta}\left\{ \begin{array}{cc}
\gamma ({r}/{L})^N &{\rm for}\ r<L \\  1 - \gamma^\prime\
{K_0(\sqrt{2} r/\xi)}&{\rm for}\  r>L\end{array}\right.\nonumber\\
\mathfrak{a}(\br)&=& {N} \hat{\mathbf{\theta}}  \left\{\begin{array}{cc}
({ r}/{ L^2}) & {\rm for}\ r<L \\
({1}/{r}) & {\rm for}\ r>L\end{array}\right.
\label{ansatz}
\end{eqnarray}
where, for continuity, $\gamma^\prime =(1-\gamma)/K_0(x)$, and for continuity of the derivative, $\gamma=x K_1(x) /[N K_0(x)+x K_1(x)]$, with
$x=(\sqrt{2}L/\xi)$.  Here the flux (charge) is uniformly distributed inside a disk of radius $L$, and is zero outside. The $K_0$ Bessel function is the solution of the linearized Landau-Ginzburg equations, so the ansatz has the correct asymptotic form at long distances from the vortex core but also (less obviously)
immediately outside it. For a given value of $\lambda/\xi$, we determine $L$ by numerically optimizing the free energy of this profile.
In the extreme Type I limit for a single vortex ($N=1$), $\xi/\lambda \gg 1$, $L \sim \lambda \log(\xi/\lambda)$ and $E_v\sim \epsilon_F/\log(\xi/\lambda)$.

Turning now to vortices in the Coulomb problem, the general features are similar and hence we can gain quantitative guidance by adopting the same variational ansatz as in Eq. \ref{ansatz}. However, in this case the ansatz does not have the proper form at long distances;  from general considerations, as first discussed in the context of abelian quantum Hall states  \cite{Sondhi:1992p1220}, it follows that in the presence of Coulomb interactions the vortex has
power-law tails in the charge and current density, which decay like $r^{-3}$ and $r^{-2}$ respectively. These do not, however, affect the basic length scales in the problem significantly.

There are, of course, features of the vortex structure that cannot be captured by our Landau-Ginzburg theory, most notably the existence of a bound Majorana zero mode and thus the
non-abelian statistics \cite{Read:2000p1015, Ivanov:2001p1366}.  While of great fundamental interest \cite{Nayak:2008p752}, these subtleties are irrelevant for present  purposes.

\noindent{\bf Vortex Binding and Optimal Droplets:} The structure of the vortex in a Type I quantum Hall fluid, with the charge density confined on a distance much shorter than that over which the order parameter is suppressed, implies that individual vortices are unstable to aggregation.  To see this, consider two vortices separated by a distance  $\xi$.  As  they share a common region of suppressed pairing ($|\Psi|< 1$) they gain an energy of order $\epsilon_0$ while paying a Coulomb energy of order $ \epsilon_0 (\lambda_C/\xi)$ which is parametrically smaller.

In the short-range case ($\lambda_C=0$) the Landau-Ginzburg theory  is formally identical to
that of a superconductor, where vortex aggregation continues indefinitely when $\lambda/\xi < 1/\sqrt{2}$ \cite{Tinkham:1996p1} and thus any number of vortices phase separate macroscopically.
 The situation is different with Coulomb interactions  where the cost of macroscopic phase separation is superextensive. Instead, when the  concentration of vortices is small, they aggregate into droplets of vortices $N_c$ and  size $\xi$. We can obtain an estimate of $N_c$  by minimizing an expression for the energy density of vortices
 $(\epsilon_{0}+ \epsilon_0 N^2\lambda_C/\xi)\overline{\delta\rho}/N$, where the expression in the bracket
 is an energy of one droplet with vorticity $N$.
In the extreme Type I limit, this leads to an estimate 
\begin{equation}
N_c \sim \sqrt{\frac{\xi}{\lambda_C}}.
\end{equation}

This estimate agrees with a variational calculation of the optimal droplet using (\ref{ansatz}). Note that a bubble with vorticity $N_c$ and charge $eN_c/4$ is, itself, a new sort of quasiparticle.  If $N_c$ is odd/even, this quasiparticle will have/not have a Majorana zero mode and corresponding non-Abelian/Abelian braiding statistics.

\noindent{\bf Phase diagram near $\nu=5/2$:}
As we move away from the ideal Pfaffian filling factor, $\nu=5/2$, we introduce a pseudomagnetic field corresponding to the difference between the average density $\bar\rho$ and the commensurate density $\rhopf$.
For short ranged interactions
($\lambda_C=0$)  the QH system behaves exactly like the
superconductor and we get phase separation between Pfaffian and metallic regions. Specifically, the additional charge
coagulates in a ``normal'' region with density $\delta \rho_c = \frac{1}{{\xi\lambda}}$
which is the transcription of the thermodynamic critical field
$H_c$ (more precisely $B_c$) to this setting. For $\bar\rho -\rhopf > \delta \rho_c$,  the entire
system is metallic.

However, in the presence of Coulomb interactions, the phase diagram of the Type I Pfaffian
is richer than that of its superconducting cousin (see Fig. \ref{fig-phasediagram}.)
Now the tendency towards phase separation is frustrated---which is a problem much studied in recent years in several contexts \cite{Jamei:2005p876,Spivak:2006p1,Ortix:2006p1}.
At small deviations from the magic density, optimal bubbles (discussed above) form a triangular lattice bubble crystal.  For larger values of $|\bar\rho -\rhopf|$, the bubble crystal typically gives way to a stripe phase.  At still larger values, one typically finds an anti-bubble or droplet phase,  which in this case consists of puddles of Pfaffian phase embedded in the majority metallic phase. The character of these various ``microemulsion phases'', and the structure of phase transitions between them will be explored in future publications.
Note that phases where the Pfaffian percolates will manifestly exhibit the QHE. In the remaining
phases the issue turns on the balance between the Pfaffian version of the proximity effect \cite{pkss-inprep}
and the effects of quantum and gauge fluctuations. The reader should contrast this complexity
with the case of the Type II Pfaffian where the density deviation is always accomodated via
a triangular lattice quasi-particle Wigner crystal forms on the background of the uniform Pfaffian state---analogous to the Abrikosov lattice in a Type II superconductor.

\noindent{\bf Concluding Remarks:} Our identification of the weakly coupled Pfaffian as a Type I QH liquid
is not particularly sensitive to various approximations, such as the neglect of fluctuations
about various saddle points, that we have employed in our treatment---essentially it depends on the existence of a regime where $\xi$ is asymptotically larger than $\lambda \sim \ell_B$. The features we obtain in that limit are macroscopic or semi-macroscopic  and thus should be robust as well.
That said, it is worth drawing attention to a subtlety that we ignored in the main text. The
metallic phase is not really immune to the pseudomagnetic field---it has ``quantum oscillations'' as a function of density (flux) reflecting the existence of various integer (and even fractional) quantum Hall states at special densities.  Sufficiently close to $\nu=5/2$ and in the strongly Type I limit, the relevant states will be fairly weak and likely to modify the properties of the bubbles only at extremely low temperatures. Of course, our considerations {\it are} sensitive to the inclusion of disorder which will destroy true long range order in the various microemulsion
phases and at sufficient strength, the underlying Pfaffian state as well.

In closing we note that the nature of the phases produced by frustrated phase separation does depend critically on the range of the frustrating interactions \cite{Spivak:2006p1}, {\it i.e. } $x$ in Eq. \ref{eq-LG}.  For $x>3$, the interactions are short-ranged, and can be lumped with the compressibility term, yielding a renormalized value of $\lambda$.  However, in the interesting dipolar case, $x=3$,  macroscopic phase separation is only marginally forbidden.
 Here, the typical density in a puddle is $\delta \rho \approx \delta \rho_c$, and the size of the bubbles defines an emergent length scale
 that grows exponentially as the size of the dipole (distance to the gate $d$) decreases. This suggests that quite generally the search for Type I QH liquids would be greatly advanced by investigating gated 2DEGs.

\noindent{\bf Acknowledgements:} We are grateful to S. Simon for helpful correspondence. BZS, SAP and SLS thank the Stanford Institute for Theoretical Physics for hospitality. This work was supported in part by NSF grants DMR-1006608 and PHY-1005429 (SAP, SLS), DMR-0758356 (SAK) and DMR-0704151 (BZS).
\bibliography{pfaffian_bib}
\end{document}